\documentclass[pre,aps,twocolumn,superscriptaddress,floatfix]{revtex4-1}
\usepackage[english]{babel}

\usepackage{color}
\usepackage{graphicx}
\usepackage{dcolumn}
\usepackage{bm}

\begin{document}


\title{
Curvature-driven AC-assisted creep dynamics of magnetic domain walls 
 }

\author{P. Domenichini}

\affiliation{ Universidad de Buenos Aires, FCEyN, Departamento de F\'{\i}sica. Buenos Aires, Argentina}
\affiliation{CONICET-Universidad de Buenos Aires, IFIBA, Buenos Aires, Argentina.}

\author{F. Paris}
\affiliation{Centro At\'omico Bariloche, CNEA, CONICET, Bariloche, Argentina}

\author{M. G. Capeluto}
\affiliation{ Universidad de Buenos Aires, FCEyN, Departamento de F\'{\i}sica. Buenos Aires, Argentina}
\affiliation{ CONICET-Universidad de Buenos Aires, IFIBA, Buenos Aires, Argentina.}

\author{M. Granada}
\affiliation{ Instituto de Nanociencia y Nanotecnología, CNEA--CONICET, Centro Atómico Bariloche, (R8402AGP) San Carlos de Bariloche, Río Negro, Argentina.}
\affiliation{ Instituto Balseiro, Universidad Nacional de Cuyo, Bariloche, Argentina}

\author{J.-M. George}
\affiliation{Unité Mixte de Physique, CNRS, Thales, Univ. Paris-Sud, Université Paris-Saclay, Palaiseau 91767, France.}

\author{G. Pasquini}
\affiliation{ Universidad de Buenos Aires, FCEyN, Departamento de F\'{\i}sica. Buenos Aires, Argentina}
\affiliation{CONICET-Universidad de Buenos Aires, IFIBA, Buenos Aires, Argentina.}

\author{A. B. Kolton}
\affiliation{Centro At\'omico Bariloche, CNEA, CONICET, Bariloche, Argentina}
\affiliation{ Instituto Balseiro, Universidad Nacional de Cuyo, Bariloche, Argentina}

\date{\today}

\begin{abstract}
The dynamics of micrometer-sized magnetic domains in ultra-thin ferromagnetic films is so dramatically slowed down by quenched disorder that the spontaneous elastic tension collapse becomes unobservable at ambient temperature.
By magneto-optical imaging we show that a weak zero-bias AC magnetic field can assist such curvature-driven collapse, making the area of a bubble to reduce at a measurable rate, in spite of the negligible effect that the same curvature has on the average creep motion driven by a comparable DC field. An analytical model explains this phenomenon quantitatively.
\end{abstract}

\pacs{Valid PACS appear here}

\maketitle

An arbitrarily weak quenched disorder has yet a notable qualitative effect in the driven motion of an extended elastic system 
such as an interface embedded in a random medium. A paradigmatic experimental 
example are pinned domain walls (DW) in thin film ``Ising-like'' ferromagnets with 
a uniform external magnetic field favouring the growth of a magnetic domain~\cite{lemerle1998domain,Metaxas2007,ferre2013universal}. 
In these materials, due to the practically unavoidable presence of random heterogeneities, 
DW velocities can vary dramatically under relatively modest changes of a weak applied field. 
Strikingly, the quantitative way the velocity assymptotically vanishes in the small field limit is {\it universal}~\cite{Ioffe1987,Nattermann1987,jeudy2016universal}, 
and is thus succesfully captured by minimal models 
that can be solved, in the limit of large systems, with 
poweful analytical~\cite{chauve2000creep} and 
numerical~\cite{kolton2006dynamics,kolton2009creep,ferrero2013numerical,ferrero2017spatiotemporal} techniques. 
These statistical-physics models yield, in particular, the celebrated creep-law 
$\ln (1/v) \propto H^{-\mu}$ for the average velocity $v$ of a 
DW in presence of weak uniform magnetic driving field $H$, 
with $\mu>0$ a universal exponent~\cite{Ioffe1987,Nattermann1987,Nattermann1990}. 
This law clearly signals the breakdown of linear-response in the collective transport. 
The success of this mathematical description unveils the basic physics 
fact that the glassy universal dynamics of DWs is mainly controlled 
by the interplay of pinning, elasticity and thermal fluctuations on the 
driven elastic interface. As such, creep theory 
is relevant for many other driven elastic systems 
with thermal or ``thermal-like'' fluctuations and quenched disorder, 
ranging from current driven vortices in superconductors~\cite{Blatter1994,nattermann2000vortex,giamarchi1998statics,giamarchi2002vortex,Kwok2016}, 
charge density waves ~\cite{brazovskii2004pinning}
to tension driven cracks~\cite{ponson2009depinning,Bonamy2011}.

\begin{figure}
\includegraphics[scale=0.24]{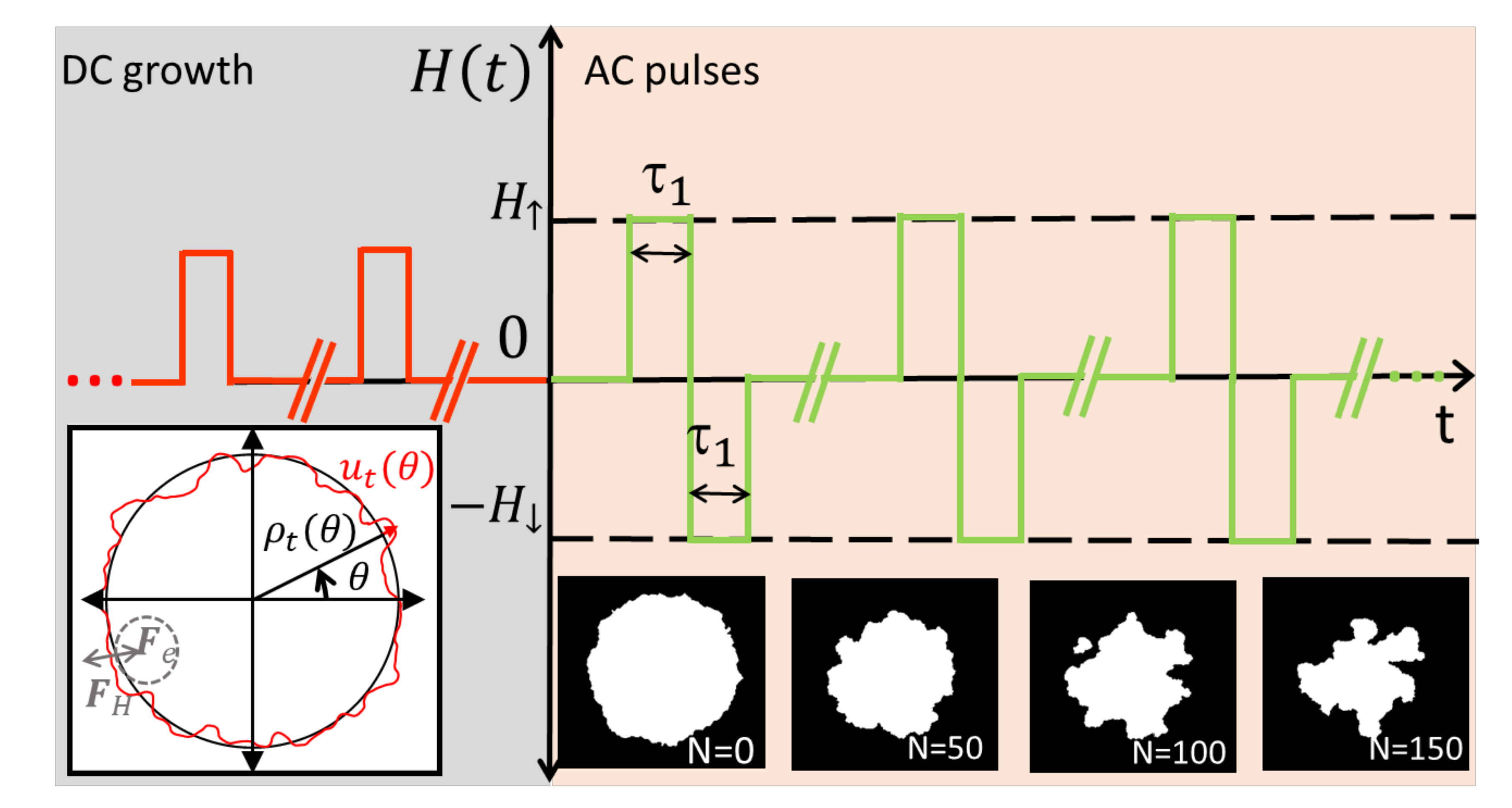}
\caption{ Schematics of the magnetic field protocol. Pulses applied to grow the domain (red) are followed by the AC-driving pulses (green) with amplitudes $H_{\uparrow}$ and $-H_{\downarrow}$. The sequence of PMOKE images in the bottom corresponds to the case $H_{\uparrow}=H_{\downarrow}$ for an increasing number  $N$ of pulses.
 Inset: Diagram of the contour 
$\Gamma_t$ of a typical domain at an arbitrary time $t$, where $u_t(\theta)$ indicates the domain profile measured from the average radius $\rho_t(\theta)$. ${\textbf{F}}_e$ and ${\textbf{F}}_H $ are the elastic and field normal forces per unit length respectively.
}
\label{fig:experiment}
\end{figure}

Many universal properties predicted by the 
creep theory, the velocity-force characteristics~\cite{Ioffe1987,Nattermann1987,Nattermann1990,chauve2000creep}, 
the rough geometry of moving DWs~\cite{kolton2006dynamics,kolton2009creep,ferrero2013numerical}, 
and even the 
event statistics behind the creep law~\cite{ferrero2017spatiotemporal,ferrero2020creep}, have been 
studied experimentally by applying external magnetic fields
\cite{lemerle1998domain,Metaxas2007,gorchon2014pinning,jeudy2016universal,pardo2017universal,
caballero2017excess,jeudy2018pinning,grassi2018intermittent,torres2019universal,Domenichini_2019} 
or external currents 
~\cite{Yamanouchi2007,Lee2011,DuttaGupta2016,Caretta2018,DiazPardo2019}
to drive DWs in ultra-thin ferromagnetic films with perpendicular anysotropy (PMA). Most of the studies focus in the DC-driven case while comparatively 
very few experimental ~\cite{Kleemann2007,Domenichini_2019} and theoretical ~\cite{Nattermann2001,Glatz2003} 
studies have focused on the {\it universal} properties that can emerge 
under a zero-bias AC-drive within the creep regime. 
Weak AC fields yield nevertheless a rich 
phenomenology which is worth studying.  
In particular, recent experiments have shown that roughly-circular 
magnetic bubbles evolve under a pure symmetric 
AC field in a very intriguing way~\cite{Domenichini_2019}.
The first interesting effect is that the (otherwise ultra-stable) 
initial bubble monotonically shrinks 
with the number of alternated positive and negative 
magnetic field pulses of equal strength, apparently ``rectifying'' 
the AC drive.
The second is the observation that the DW
roughness increases at a much faster rate in the 
AC protocol compared to the DC for the same amplitude 
of the drive. 
An example of such evolution, captured by successive MO images, is shown in Fig. \ref{fig:experiment}. These two intriguing effects have not been explained yet.

In this Letter we show that the 
elastic pressure arising from the domain mean curvature, 
even being orders of magnitude weaker than the driving field pressure, 
is the responsible for the 
shrinking of the domain area under AC fields. 
To show this, we first derive a model for the 
AC-driven DW dynamics and second, 
we quantitatively test two of its predictions experimentally: 
(i) the pulse asymmetry needed to stabilize the average size of the ``beating domain'' 
and, 
(ii)  The area collapse rate in the initial dynamics for the case of symmetric positive-negative pulses. 
Finally, a qualitative argument is given to explain the AC enhancement of 
the DW dynamic roughening and its effect on the area collapse dynamics.


The proposed model is very simple. Let us consider a segment of a DW 
in a film of thickness $d$. In the absence of disorder and thermal fluctuations, the total force $\bf F$ per unit-length acting on the effectively one-dimensional DW is  the sum of the magnetic field force ${\bf F}_H=2 M_s d H \hat n$ and the elastic force ${\bf F}_e=\sigma d \kappa \hat n$, 
being $M_s$  the saturation magnetization, $\sigma$ the DW surface tension, $\kappa$ the DW signed local curvature 
~\footnote{We are implicitly assuming smooth enough DW distortions such that the DW width is $w \ll 1/\kappa$}.
Both forces are parallel to the DW local outward normal $\hat n$ (see inset in Fig. \ref{fig:experiment}).
For instance, for a perfectly circular bubble of radius $R$,
the ``Young-Laplace'' effective magnetic-field $C \kappa \approx -C/R$, with the constant $C = \sigma/2 M_s$, 
opposes domain growth.
Assuming an overdamped dynamics we have $\dot R = m(H-C/R)$, 
with $m$ the effective DW mobility. 
For $H=0$ this predicts the linear decay 
$A_t = A_0-(2\pi C m) t$, 
where $A_t \equiv \pi R_t^2$ is 
the shrinking circle area. 
The bubble is hence unstable and its 
lifetime scales as $\sim A_0$. This result 
can be also obtained from the 
general Allen-Cahn equation 
~\cite{bray1994}. Quite remarkably 
the result actually holds for {\it any} simple 
time dependent
closed curve ${\Gamma_t}$
~\cite{white2002evolution} for which we can write 
\begin{eqnarray}
\frac{dA_t}{dt} &=& \int_{\Gamma_t} v_t({\bf r}_s) ds
\label{eq:dAdt}
\end{eqnarray}
with $v_t({\bf r}_s)$ the local instantaneous normal 
velocity at point ${\bf r}_s$ in $\Gamma_t$. 
If we assume a linear and instantaneous response 
$v_t({\bf r}_s)=m(H_t+C\kappa_t({\bf r}_s))$, 
with $\kappa_t({\bf r}_s)$ the instantaneous 
signed local curvature we easily obtain, 
using the topological index of the curve  
$\int_{\Gamma_t} \kappa_t({\bf r}_s)ds=-2\pi $,  
the rate $\frac{dA_t}{dt}=m H_t P_t - 2\pi C m$, 
with $P_t\equiv \int_{\Gamma_t} ds$ the perimeter. 
This generalizes the $H_t=0$ constant decay 
rate $\frac{dA_t}{dt}=-2\pi C m$ to {\it any} 
initial simple closed curve $\Gamma_t$.

In the presence of quenched and thermal 
disorder the above results are not valid as the 
normal velocity $v_t({\bf r}_s)$ is 
in general expected to be an inhomogeneous non-linear 
function of $H_t+C\kappa_t$. 
Assuming again instantaneous response 
we can write Eq.(\ref{eq:dAdt}) as
\begin{equation}
\frac{dA}{dt} 
\approx \int_{\Gamma_t} {V}_T(H_t+C\kappa_t({\bf r}_s),{\bf r}_s) ds 
\label{eq:dAdtm}
\end{equation}
with $V_T(h,{\bf r})$ a temperature
and position dependent velocity response to 
a local field $h_t=H_t+C\kappa_t$. 
We will argue that for weak 
enough fields, $V_T(h,{\bf r})$ in Eq.~(\ref{eq:dAdtm}) can 
be approximated by 
the well known creep law for DC-driven DWs.
This hydrodynamic approach can be formally justified: in the creep regime, DW velocity is mainly  controlled by creep events with a cut-off radius 
estimated to be, for ultra-thin ferromagnet, less than $ 0.1 \mu m$ \cite{ferrero2017spatiotemporal,grassi2018intermittent}, clearly well below the $\sim 1\mu \text{m}$ PMOKE resolution. 
Therefore, larger size fluctuations are expected to introduce only negligible logarithmic corrections ~\cite{ferrero2013numerical} into the 
creep law 
$V_T(h) \sim \exp[-(T_d/T)(H_d/h)^\mu]$ 
that describes the DW velocity 
in terms of 
the effective field $h$, temperature $T$ and also the disorder and elasticity through $T_d$, 
and $H_d$ \footnote{See Ref.\onlinecite{jeudy2018} for values of $H_d$, $T_d$ and also the associated fundamental Larkin length $L_d$ in different magnetic materials.}. 
Similarly, the 
characteristic time associated to individual 
creep events is much smaller than 
the experimental time-scale 
used for resolving DW displacements so
the velocity response can be considered local and  instantaneous. 

Replacing the creep velocity $V_T(h_t,{\bf r})$ 
in Eq.~(\ref{eq:dAdtm}) is a step forward but still yields 
a non-closed equation for $dA_t/dt$ as it requires 
the knowledge of the time dependent curvature field 
$\kappa_t({\bf r})$, together with a model for 
the spatially fluctuating pinning parameters 
of the creep law. 
Nevertheless, to extract the basic physics 
some progress can be made by first making 
the well justified approximation that $H_t \gg C\kappa_t$ 
\footnote{
Typical external fields $H_t$ applied in creep experiments induce DW motion  that can be measured by PMOKE, whereas for $H_t=0$, DW motion solely driven by the curvature effective field $C\kappa_t$ is hardly observed in typical experimental time-scales and micrometer-sized  domains, so for typical finite fields $H_t\gg C\kappa_t$.
See \cite{SuppMat} for more details}.
Second, the complexity of Eq.(\ref{eq:dAdtm})
is greatly reduced if we neglect the heterogeneity of the creep-law 
and replace it by its average 
$V_T(h,{\bf r}) \approx V_T(h)$ or 
velocity-field characteristics. 
This approximation is not equivalent 
to neglect disorder completely, as $V_T(h)$  
is in general quite different from the $V_T(h)\propto h$ expected for an homogeneous sample, particularly in the strongly nonlinear creep-regime. 
Developing then at first order in $C\kappa_t$ 
from Eq.(\ref{eq:dAdtm}) we obtain 
\begin{equation}
\frac{dA_t}{dt} 
\approx  {V}_T(H_t) P_t - 2 \pi C {V'_T}(H_t), 
\label{eq:dAdt1st}
\end{equation}
only relating the geometric variables $A_t$ and $P_t$. 
The position and time dependent curvature $\kappa_t({\bf r})$ 
disappears thanks to the topological 
invariant $\int_{\Gamma_t} \kappa_t=-2\pi$~\cite{SuppMat}. 
Let us now focus on the experiments and make some concrete 
predictions with Eq.(\ref{eq:dAdt1st}). 

Our measurements were carried out in ultrathin ferromagnetic films with PMA, by Magneto-optical imaging, using a homemade polar Magneto-optical Kerr effect (PMOKE) microscope. Two kinds of  samples from different sources were used:  a Pt/Co/Pt magnetic monolayer (S1) and a Pt/[Co/Ni]4/Al multilayer (S2), both  grown by DC magnetron sputtering \cite{Rojas2016, Quinteros2020}.
{Helmholtz coils allow to apply well conformed square magnetic field pulses with amplitude $H$ up to $700$ Oe and duration $\tau_1 > 1$ ms.}
{DW dynamics is characterized with the usual quasistatic technique (see \cite{SuppMat} for experimental details). The AC field is applied to an already grown domain (see Fig. 
\ref{fig:experiment}) and consists
in alternated square pulses of 
identical duration $\tau_1$ and amplitude $H=H_{\uparrow}>0$ (expanding the domain), 
and $H=-H_{\downarrow}<0$ (compressing the domain). The two pulses are periodically
repeated with period $\tau \geq 2\tau_1$, as schematized in Fig. 
\ref{fig:experiment}. The magnitudes of all applied fields are such that 
the creep-law with $\mu=1/4$ is well observed 
in the DC protocol (see \cite{SuppMat}).} 

Since we are only interested in the smooth evolution 
of $A_t$ and $P_t$ as a function of the number $N$ of AC cycles 
we define ${\cal A}$ and ${\cal P}$ 
such that $d{\cal A}/dN \equiv A_{n+1}-A_{n}$ and 
$d{\cal P}/dN \equiv P_{n+1}-P_{n}$, where $n\equiv t/\tau$. 
Integrating Eq.(\ref{eq:dAdt1st}) from $t=n\tau$ to $t=(n+1)\tau$
we thus obtain
\begin{equation}
\frac{d{\cal A}}{dN} \approx 
- 2\pi CV'_T(H_{\uparrow})\tau -  
\frac{\tau}{4}V_T(H_{\uparrow}) \frac{d{\cal P}}{dN}
+ \frac{\tau}{2}\Delta H V'_T(H_{\uparrow}) {\cal P},
\label{eq:prediction}
\end{equation}
where $\Delta H=H_{\uparrow}-H_{\downarrow} \ll H_{\uparrow}$ 
quantifies a possible pulse asymmetry, and we have 
used the expected symmetry $V_T(h)=-V_T(-h)$. 

\begin{figure}
\includegraphics[scale=0.45]{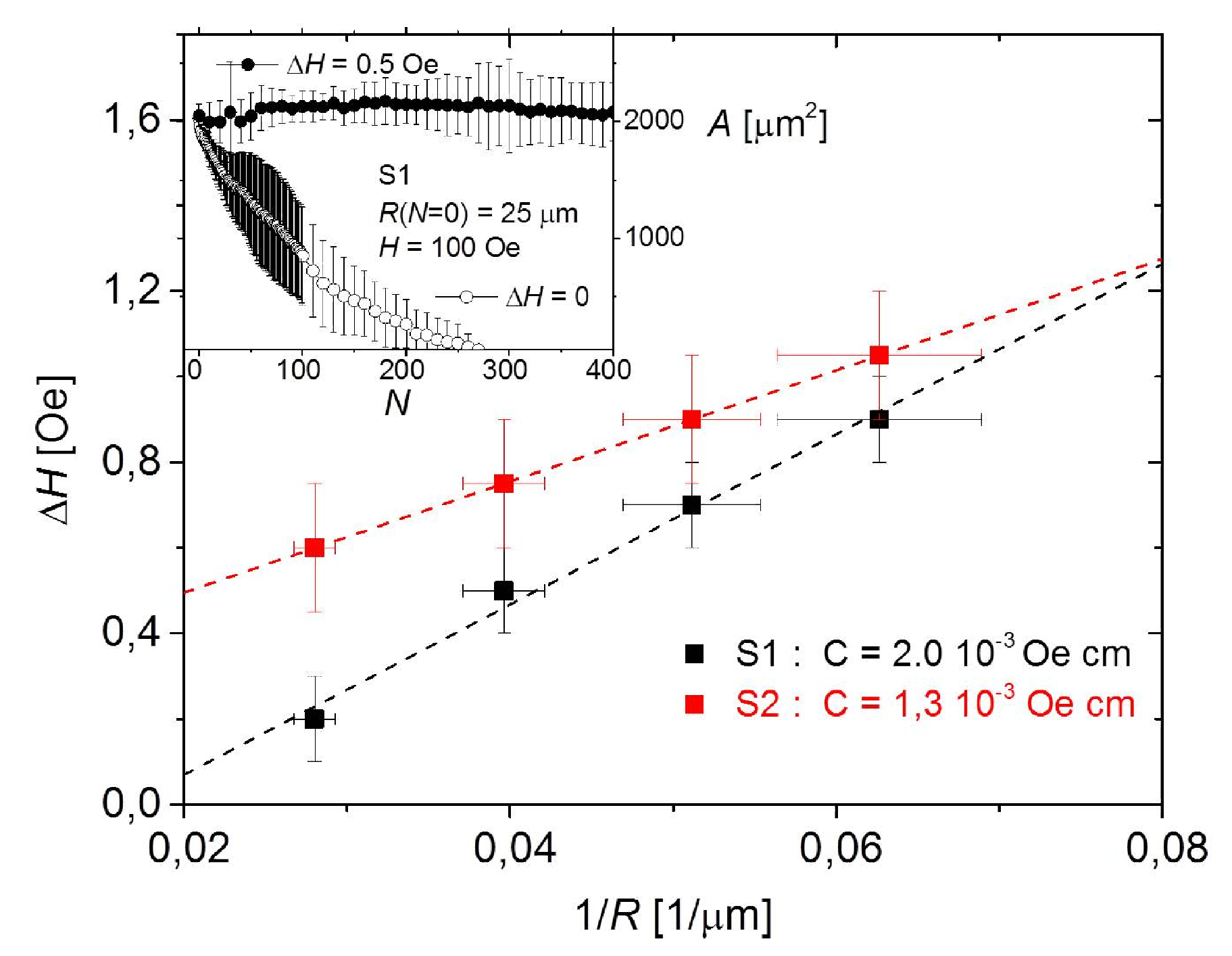}

\caption{Field pulse asymmetry $\Delta H$ used to compensate the curvature collapse of domains with different initial radii, $R$. Inset: Evolution of the domain area $A$ with the pulse number, for symmetric pulses $\Delta H = 0$ and for nearly compensating asymmetric pulses $\Delta H = 0.5\;\text{Oe}$. 
}

\label{fig:compensation}
\end{figure}

We test Eq.(\ref{eq:prediction}) in two different ways. On one hand, we can choose 
$\Delta H=\Delta H^*$ such that ${d{\cal A}}/{dN}={d{\cal P}}/{dN}=0$,
\begin{eqnarray}
\Delta H^*=\frac{2C}{{\cal R}}
\label{eq:compensate}
\end{eqnarray}
where we have defined ${\cal R}\equiv {\cal P}/2\pi$, 
approximately the observed average domain radius.
This is a simple but rather general prediction: $\Delta H^*$ is independent of $V_T$ only provided that $V_T(h)=-V_T(-h)$, and
of the AC parameters $\tau$ and $H_{\uparrow}$.
In physical terms, Eq.(\ref{eq:compensate}) states that even weak compressing forces arising from mean curvature are relevant because they 
{\it break the forward-backward symmetry} of the DW velocity in the AC field.
Importantly, Eq.(\ref{eq:compensate}) connects 
with micro-magnetism through $C=\sigma/2M_s$. 
Using that ${\cal A}_N \approx A_N$,  in Fig. \ref{fig:compensation} 
we test Eq.(\ref{eq:compensate}) experimentally. The  main panel 
shows the field asymmetry $\Delta H $ stabilizing the average area $A$ 
of initially nucleated domains 
with different initial radius 
$R$. An example of such compensation is shown in the inset, 
where the evolution of ${\cal A}(N)$ under symmetric field pulses and asymmetric compensating pulses 
are compared for sample S1, with an initial $R= 25 \mu$m. The corresponding videos 
are available in the Supplemental Material. 
For both samples,
there is a good agreement with 
the linear relation between $\Delta H^*$ and $1/{\cal R}$ 
predicted in Eq.(\ref{eq:compensate}), for
four $R$ ranging from $15\mu m$ to $35\mu m$. 
The ordinates 
predicted to be zero in Eq.(\ref{eq:compensate}) 
are small for the two samples, 
compatible with a  small hazard DC field present in the Lab.
The fitted value of $C$ is in both cases of 
order $10^{-3} \text{Oe} \;\text{cm}$, fairly agreeing 
with $C_{S1}= 2.1 \times 10^{-3}$ Oe cm and $C_{S2}=1.2 \times 10^{-3}$ Oe cm
estimated as $C=\sigma/2M_s$ from the 
micromagnetic parameters 
respectively \cite{SuppMat}.

\begin{figure} 
\includegraphics [scale=0.38] {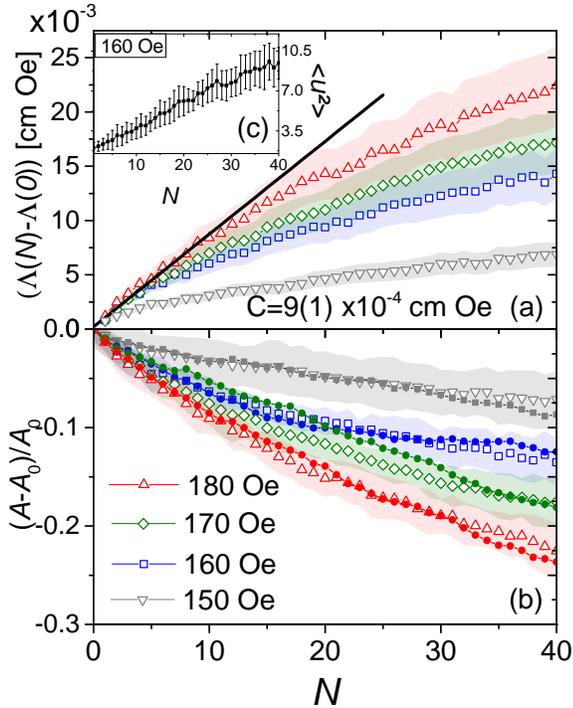}
\caption{Open symbols show the evolution of $\Lambda(N)$ (see text) (a) and of the relative domain area change $\Delta A/A_0$ (b), for different amplitudes of symmetric pulses in sample S2. 
The black-line in (a) is a linear fit of the initial evolution of $\Lambda(N)$, yielding the micromagnetic constant $C$. In (b) numerical simulation results are shown with full symbols. Inset (c): DW mean squared displacements for $H=160$ Oe pulses.}

\label{fig:areadecay}
\end{figure}

Let us now go further and 
focus in the interesting case of symmetric field pulses 
$H_{\uparrow}=H_{\downarrow}=H$, i.e. $\Delta H=0$, 
From Eq.(\ref{eq:prediction}) we 
simply predict 
\begin{equation}
-\frac{d}{dN} \left[\frac{{\cal A}+\frac{\tau}{4}V_T(H){\cal P}}{2\pi V'_T(H)\tau}\right] = \frac{d{\Lambda}}{dN} \approx C.
\label{eq:areadecay}
\end{equation}
For circular DWs with radius $R_t$, this equation 
can be readily obtained from $dR_t/dt \approx V(H_t + C\kappa_t)$ 
with $\kappa_t=-1/R_t$.
Remarkably however, Eq.(\ref{eq:areadecay}) is valid regardless of the circular shape assumption (see ~\cite{SuppMat} for further details) and contains the spontaneous ($H=0$) collapse as a special case.
Fig. \ref{fig:areadecay}(a) shows the evolution of the function 
$\Lambda (N)$ defined in Eq.(\ref{eq:areadecay}), for 4 different field amplitudes, measured in sample S2.
The initial slope for the highest amplitudes,  using the creep-regime velocity-field characteristics measured in S2,  gives
$C\approx 10^{-3}\text{Oe}\;\text{cm}$, again in fair agreement with the 
micromagnetic estimate for $C$, hence reinforcing the curvature argument. 

Results shown in Figs. \ref{fig:compensation}
and \ref{fig:areadecay}(a) 
confirm that surface tension forces arising from curvature
are responsible for the domain collapse, in spite of being two orders of magnitude smaller than the AC forces. They also validate the proposed model but, 
as can be appreciated in Fig. \ref{fig:areadecay}(a), only for 
the very first few AC cycles (small $N$); the lowest the field the sooner the deviation. We argue that these deviations are due to large-scale dynamic roughening, neglected in our simple model. 
To show it we exploit that creep dynamics displays a ``depinning-like'' regime 
upon coarse-graining many creep events~\cite{chauve2000creep,kolton2006dynamics,kolton2009creep} 
and thus numerically emulate the experimental 
protocol using the time-dependent Ginzburg-Landau model
~\cite{Caballero2018,Caballero2020b,guruciaga2019ginzburglandau,caballero2020degradation},
\begin{equation}
\eta \partial_t \phi = c\nabla^2 \phi + 
\epsilon_0[(1+r(x,y)) \phi - \phi^3] + \tilde{h}_t,
\label{eq:phi4}
\end{equation}
near the depinning transition, where $\phi\equiv \phi(x,y,t)$ models the local 
magnetization, and $r(x,y)$ 
an uncorrelated 
random-bond type of disorder 
of strength $r_0$ 
\footnote{Note that we do not attempt to simulate creep but the 
``depinning-like'' dynamics which effectively emerges by coarse-graining many creep events~\cite{chauve2000creep}. 
Although the time-dependent 
Ginzburg-Landau model with 
thermal noise has all the 
ingredients, the universal creep-formula that we can observe experimentally has 
been so far numerically tested only in much 
simpler models which can be solved with algorithms able to overcome the ultra-slow dynamics and high futility that characterize the deep universal creep regime~\cite{kolton2009creep}}. 
The {\it effective} AC field 
$\tilde{h}_t = \pm {\tilde h}_0$ of period ${\tilde \tau}$
is chosen so to impose a given average displacement of the 
DW in half a period comparable to that observed in the experiments.
After nucleating a circular domain 
with saturation magnetization $\phi_s \sim 1$, 
this model generates a closed curve $\Gamma_t$ 
(i.e. $\phi({\bf r}_s)=0$ for ${\bf r}_s \in \Gamma_t$) 
describing a DW with width $\delta \sim \sqrt{c/\epsilon_0}$,
and surface tension $\sigma \sim \sqrt{c \epsilon_0}$, 
driven by an AC field in a disordered environment. When $r_0=0$, Eq.(\ref{eq:areadecay}) is very accurately satisfied if we replace  $V_T(H)\tau \to {\tilde h}_0 {\tilde \tau} \delta/\eta$ 
and $V'_T(H)\tau \to {\tilde \tau} \delta/\eta$ and $C \to \sqrt{c\epsilon_0}$~\cite{SuppMat}. 
When $r_0\neq 0$ however, we go beyond the homogeneity assumption and a deviation from Eq.(\ref{eq:areadecay}) similar to the experiment is found. Strikingly, we can empirically adjust $r_0$, and then tune ${\tilde h}$ 
and ${\tilde \tau}$ so to accurately reproduce the experimental data for the different fields, as shown in Fig.\ref{fig:areadecay}(b). 
Large-scale dynamic roughening within the creep regime hence slow-down the curvature-driven collapse. The better agreement for small $N$ between the prediction (Eq.(\ref{eq:areadecay})) and the experiment (Fig.\ref{fig:areadecay}(a)) for increasing AC-fields is explained by the smoothing effect of the DW velocity.

Finally, we aim to explain
why the AC dynamic roughening discussed above is enhanced with respect to the DC-driven case~\cite{Domenichini_2019}.
As a general fact, we expect that a recently nucleated driven DW
will display, as it correlates with the disorder, a growing DW mean squared width, $w_t^2 \sim t^{2\zeta/z}$, with 
$\zeta$ and $z$ the roughness and dynamic exponents, 
respectively~\citep{BarabasiBook}. If the large-scale geometry of a DW in 
the DC creep regime is described by the 
Edwards-Wilkinson (EW)
equation with an effective temperature
~\cite{chauve2000creep,kolton2009creep,grassi2018} 
we expect $z=2$ and $\zeta=1/2$, so $w_t^2 \sim t^{1/2}$ for the 
1d DW. For the AC case we 
expect instead a temporally correlated noise since the DW can revisit repetitively the same disorder in its oscillatory motion. 
If we model such colored noise $\eta(x,t)$ 
with an exponent $\psi>0$, such that 
for two points in a DW segment
$\langle \eta(x,t)\eta(x,t')\rangle \sim \delta(x-x')|t-t'|^{2\psi-1}$ ($\psi=0$ for uncorrelated noise)
linear theory predicts $z=2$ and $\zeta=1/2+2\psi > 1/2$~\cite{BarabasiBook}. 
Therefore $w^2_t$ should grow {\it faster} than in the DC case. To test this idea we describe the experimental DW in polar coordinates $\rho(\theta,t)$
\footnote{For small number 
of cycles $N$, the DW position is found to be uni-valued in polar coordinates.}, and define 
$u_t(\theta)=\rho_t(\theta)-R_t$ with 
$R_t\equiv \langle \rho_t(\theta)\rangle$ the 
angle-averaged radius (see inset in Fig. 1), and then compute 
$w^2_t = \langle u_t^2 \rangle$. 
The inset of Fig.\ref{fig:areadecay} 
shows an example of the AC evolution of $w^2_N$ in sample S2. 
Interestingly, it can be seen that $w^2_N \sim N$, 
faster than the prediction for uncorrelated noise and compatible with 
correlated noise with $\psi \approx 1/2$ in the relevant $N$-range~\footnote{Since the oscillatory motion is superimposed with a slow drift we can not discard a crossover at large $N$ towards the Edwards-Wilkinson growth.}. 
This may explain qualitatively why 
DWs in the AC protocol are rougher than in the DC protocol for identical field amplitude and 
time window, as observed experimentally~\cite{Domenichini_2019}, and in recent simulations~\cite{caballero2020degradation}.

Summarizing, we have proposed and experimentally tested a 
model for the DW creep dynamics of an isolated magnetic domain in an  
ultra-thin ferromagnet under  
AC fields at ambient temperature. 
We showed that curvature effects, with a negligible effect on the 
average DC-driven motion, play nevertheless a central role in the 
AC-driven case. The intriguing ``rectification effect'' in the zero-bias AC case of Ref.\onlinecite{Domenichini_2019} is 
then explained by the curvature-induced symmetry breaking of forward-backward  
DW motions. Rather strikingly, the same curvature effects are unable to produce, without AC-assistance, any experimentally observable DW displacement~\footnote{See Ref.~\onlinecite{SuppMat} Sec VIII for estimates of domain collapse lifetimes}. 
We have also explained, qualitatively, the 
AC enhancement of large-scale dynamic roughening.
Although we have focused in their important role in the AC-assisted motion, curvature effects can be relevant in some DC-driven systems as well: a non-steady velocity in 
DC-driven circular domains was experimentally reported~\cite{moon2011}; 
on the other hand, in thin and narrow ferromagnetic wires the universal creep-law 
is satisfied by the DC-driven steady
DW velocity only if
an effective ``counterfield'' $\Delta H$~\cite{diez2018,herrera2020},
proportional to the observed average curvature of the narrow DW, is added, in agreement with our arguments. In the latter case, however, average curvature is not inherited from the initial conditions but steadily maintained by the strong localized ``dynamic friction'' at the wire edges. Due to the simplicity and generality of our arguments, we hope that the present work will open new perspectives for modelling and controlling DW creep motion in a variety of elastic systems, far beyond ferromagnetic  films.  
\begin{acknowledgements}
We specially thank J. Curiale, G. Durin, E. Ferrero, V. Jeudy and A. Rosso for useful discussions,  This work was partially supported by Consejo Nacional de Investigaciones Científicas y Técnicas - Argentina (CONICET), the University of Buenos Aires and
grants PICT2016-0069 (MinCyT) and UNCuyo2019-06/C578.
\end{acknowledgements}

\bibliography{referencias,kolton}

\end{document}